\documentclass[12pt]{article} 
 1                                        
\def\vec#1{\mbox{\protect\boldmath $ #1 $}}                                
                               
\begin{document}                                                              
\begin{center}                                                                
{ \large\bf SUBTLETIES IN CPT-TRANSFORMATION FOR MAJORANA FERMIONS \\}
\vskip 2cm  
Marek Nowakowski \\                                
Universidad de los Andes, Departamento de Fisica, A.A. 4976,\\ Santafe de Bogota D.C., Colombia \\
and \\
Instituto de Fisica, Universidad de Guanajuato, Apdo Postal E-143, 
37150 Leon, Gto, Mexico
\end{center}
\vskip 2cm                             
\begin{abstract}
We point out the relevance of the so-called Majorana creation phase in the $s$-channel matrix 
elements in connection with the CPT-transformation of the latter.
\end{abstract}     
\newpage
By now Majorana particles play an important role in particle models, either by invoking supersymmetry
\cite{susy}
or by speculating about Majorana nature of massive neutrinos \cite{petcov, kaiser, mohapatra}. 
It is known that Majorana fermions are
defined up to a phase $\lambda_M$, often called ``creation phase''. This phase is conventional
and therefore physical results should not depend upon it. It is, however, appreciated that 
in many situations the most 
convenient choice of this phase is often not $\pm 1$ \cite{kaiser}. 
Once these phases are fixed, they will enter
the individual expressions of the coupling constants of the Majorana particles albeit, as stated 
above, the final physical results are independent of the choice of $\lambda_M$'s. In this short note 
we point out that there exist yet another place where the creation phase makes it appearance.
This is in connection with $s$-channel matrix elements where an overall dependence of the creation phase
has to be taken into account. The negligence of this $\lambda_M$ dependence of such matrix elements
leads to contradictions with CPT-invariance, as shown below.
This is especially relevant when we parameterize, 
in principle, unknown matrix elements.

Let us start by introducing some definitions. The Majorana field $\Psi_M$ is defined be demanding 
that $\Psi_M$ be self-conjugate, up to a phase, i.e.
\begin{equation} \label{definition}
\lambda_C C \overline{\Psi}_M^T(\vec{x},t)
=\lambda_C\Psi_M^C(\vec{x},t)
\nonumber \\
=\lambda_M'\Psi_M(\vec{x},t)
\end{equation}
where $C$ is the charge conjugation  matrix satisfying $C^{-1}=C^{\dagger}=C^T=-C$ and 
$\lambda_C$ as well as $\lambda_M'$ are arbitrary phases. In Fock space we can write
\begin{equation} \label{fock}
\Psi(\vec{x},t)=\int [dk] \sum_{\lambda}\left[a(\vec{k},\lambda)u(\vec{k},\lambda)
e^{-ikx}+\lambda_Ma^{\dagger}(\vec{k},\lambda)v(\vec{k},\lambda)e^{ikx}\right]
\end{equation}
where $[dk]$ is a three dimensional integration measure depending of the normalization of 
the spinors, $\lambda$'s are particle's helicities and $\lambda_M \equiv -\lambda_C\lambda_M'^{\ast}$.
The latter is known as the Majorana creation phase. That it can be chosen at will can be best seen
at the place in the langrangian where Majorana fields are defined. For instance,  
a Majorana mass term for neutrino of the form 
$(-1/2)\overline{\nu_{iL}^c}M_{ij}\nu_{jL}$ can be diagonalized
by an unitary matrix $U$ with $M=(U^{\dagger})^TmU^{\dagger}$ such that $m={\rm diag}(m_i, m_j,...)$.
To obtain positive definite masses $|m_i|$ the possible phases in the parameters $m_i$ can be absorbed
in two ways. One way is to absorb the phases  as creation phases in the definition of the Majorana
fields. The other is simply to redefine the rotation matrix as $U'^{\dagger}=SU^{\dagger}$
with $S$ being a phase diagonal matrix \cite{petcov}.  
This, of course, also means that we are allowed to 
arbitrarily introduce phases in the definition of the Majorana fields which read in general as
\begin{equation} \label{majorana}
n_i=n_{iL}-\lambda_M^{(i)}(n_{iL})^C
\end{equation}
with $n_{iL}=U'^{\dagger}_{ij}\nu_{jL}$ and arbitrary phases $\lambda_M^{(i)}$. It can be shown that
$\lambda_M'$ in equation (\ref{definition}) has to be real i.e. $\pm 1$. As a result we have 
$\lambda_M^{(i)}=\pm \lambda_C^{(i)}$.

Let us now consider the electromagnetic matrix element of the current $j^{\mu}_{em}$ sandwiched
between one particle Majorana states $\phi_{(\vec{k}_i, \lambda_i)}=a^{\dagger}(\vec{k}_i,
\lambda_i)\Omega$, $\Omega$ being the vacuum state.
\begin{equation} \label{tchannel}
\left(\phi_{(\vec{k}_1, \lambda_1)}, j^{\mu}_{em}(x)\phi_{(\vec{k}_2,\lambda_2)}\right)
\nonumber \\
= e^{-i(k_1-k_2)x} \bar{u}(\vec{k}_1, \lambda_1)\Sigma^{\mu}_t(k_1, k_2)u(\vec{k}_2, \lambda_2)
\end{equation}
where the index $t$ reminds us that we are in $t$-channel matrix elements (i.e. one outgoing and one 
incoming Majorana particle).
$\Sigma^{\mu}_t$ can be as 
usual decomposed into form-factors \cite{mohapatra} in agreement with Lorentz covariance, hermiticity of $j_{em}^{\mu}$
and gauge invariance, viz.
\begin{eqnarray} \label{formfactor}
\bar{u}(\vec{k}_1 , \lambda_1)\Sigma^{\mu}_t(k_1, k_2)u(\vec{k}_2, \lambda_2)
&=&\bar{u}(\vec{k}_1,\lambda_1)\biggl [\gamma^{\mu}F_1(t) + i\sigma^{\mu \nu}q_{\nu}F_2(t)\nonumber \\
+i\epsilon^{\mu \nu \alpha \beta} \sigma_{\alpha \beta}q_{\nu}F_3(t)
&+&(q^{\mu}-{t \over 2m}\gamma^{\mu})\gamma_5 F_4(t)\biggr]u(\vec{k}_2, \lambda_2)
\end{eqnarray}
with the standard meaning of the from-factors at $q^2=t=0$.

It is known that for Majorana fields $\overline{\Psi}_M\gamma_{\mu}\Psi_M=\overline{\Psi}_M
\sigma_{\mu \nu}\Psi_M=0$ and thus the only non-zero from-factor for Majorana particles is
the anapole form-factor $F_4$. Interestingly, one can also exclude $F_{i=1,2,3}$ for Majorana 
states by imposing CPT-invariance of the electromagnetic matrix element 
\cite{cpt1, cpt2, cpt3, cpt4,cpt5}. 
For this purpose we need
the charge conjugation,  parity and time-reversal transformations
acting on a mass-eigenstate Majorana field in Fock space
\begin{eqnarray} \label{PT}
\vec{{\cal C}}\Psi_M(\vec{x},t)\vec{{\cal C}}^{\dagger} 
&=&\lambda_M'\Psi_M(\vec{x},t)
\nonumber \\
\vec{{\cal P}}\Psi_M(\vec{x}, t)\vec{{\cal P}}^{\dagger}&=&\lambda_P\gamma^0\Psi_M(-\vec{x}, t)
\nonumber \\
\vec{{\cal T}}\Psi_M(\vec{x}, t)\vec{{\cal T}}^{\dagger}
&=&\lambda_T\gamma_5 C\Psi_M(\vec{x}, -t)
\end{eqnarray}
It is straightforward to deduce from that the combined transformation $\vec{{\cal S}}\equiv
\vec{{\cal C}}\vec{{\cal P}}\vec{{\cal T}}$ for the annihilation and creation operators, respectively
\begin{eqnarray} \label{cpt}
\vec{{\cal S}}a(\vec{k},\lambda)\vec{{\cal S}}^{\dagger}&=&-
\lambda_S\lambda_M^{\ast}(-1)^{{-\lambda -1 \over
2}}a(\vec{k}, -\lambda)
\nonumber \\
\vec{{\cal S}}a^{\dagger}(\vec{k},\lambda)\vec{{\cal S}}^{\dagger}&=&
+\lambda_S\lambda_M(-1)^{{-\lambda -1 \over
2}}a^{\dagger}(\vec{k}, -\lambda)
\end{eqnarray}
in which $\lambda_S=\lambda_C\lambda_P\lambda_T=\pm i$. Applying now the CPT-transformation to 
the matrix element (\ref{tchannel}) gives 
\begin{eqnarray} \label{ttrafo}
&&\bar{u}(\vec{k}_1 , \lambda_1)\Sigma^{\mu}_t(k_1, k_2)u(\vec{k}_2, \lambda_2)
\nonumber \\
&=&\left(\Omega a^{\dagger}(\vec{k}_1,\lambda_1), j^{\mu}_{em}(0)a^{\dagger}(\vec{k}_2, \lambda_2)
\Omega \right)
\nonumber \\
&=&\left(\Omega, \vec{{\cal S}}^{\dagger}\vec{{\cal S}}a(\vec{k}_1,\lambda_1)
\vec{{\cal S}}^{\dagger}\vec{{\cal S}}j^{\mu}_{em}(0)\vec{{\cal S}}^{\dagger}\vec{{\cal S}}
a^{\dagger}(\vec{k}_2, \lambda_2)\vec{{\cal S}}^{\dagger}\vec{{\cal S}}\Omega\right)
\nonumber \\
&=& (-1)^{{-\lambda_1 -\lambda_2 \over 2}}\left(\Omega, a(\vec{k}_1, -\lambda_1)j^{\mu}_{em}(0)
a^{\dagger}(\vec{k}_2, -\lambda_2)\Omega\right)^{\ast}
\nonumber \\
&=& (-1)^{{-\lambda_1-\lambda_2 \over 2}}
\biggl[
\bar{u}(\vec{k}_1, -\lambda_1)\Sigma^{\mu}_t(k_1, k_2)u(\vec{k}_2, -\lambda_2)\biggr]^{\dagger}
\end{eqnarray}
In equation (\ref{ttrafo}) we have used $\vec{{\cal S}}j^{\mu}_{em}(0)
\vec{{\cal S}}^{\dagger}=-j^{\mu}_{em}(0)$, $\lambda_S^2=-1$ and $\vec{{\cal S}}\Omega=\Omega$.
The details of the calculation in (\ref{ttrafo}) follow essentially the steps given in
\cite{cpt1, cpt2, cpt3, cpt4, cpt5}.
It is now convenient to introduce the following notation for the linear independent $\gamma$-matrices
which we will in general denote by $\Gamma$
\begin{eqnarray} \label{gamma}
\Gamma^{\dagger}&=& \eta_0[\Gamma]\gamma^0\Gamma \gamma^0
\nonumber \\
\Gamma^T &=& \eta_C[\Gamma]C\Gamma C^{-1} \nonumber \\
\gamma_5\Gamma \gamma_5 &=& \eta_5[\Gamma]\Gamma
\end{eqnarray}
where the $\eta$'s are pure signs depending on the matrix $\Gamma$. We can now calculate the last
last expression in (\ref{ttrafo}) for an arbitrary $\Gamma$ matrix. The result reads
\begin{eqnarray} \label{ttrafo2}
&& (-1)^{{-\lambda_1-\lambda_2 \over 2}}
\left[
\bar{u}(\vec{k}_1, -\lambda_1)
\left\{
\begin{array}{c}
 1 \\
i 
\end{array}
\right\}
\Gamma u(\vec{k}_2, -\lambda_2)\right]^{\dagger}
\nonumber \\
&=& \eta_0[\Gamma]\eta_C[\Gamma]\eta_5[\Gamma]\bar{u}(\vec{k}_1, \lambda_1)
\left\{
\begin{array}{c}
 -1  \\
i 
\end{array}
\right\}\Gamma u(\vec{k}_2, \lambda_2)
\end{eqnarray} 
Since $\eta_0[\Gamma]\eta_C[\Gamma]\eta_5[\Gamma]=+1,-1,-1,-1$ for $\Gamma=\gamma_{\mu},
\sigma_{\mu \nu},\gamma_5, \gamma_{\mu}\gamma_5$, respectively, we conclude that the only surviving
electromagnetic form-factor in (\ref{formfactor}) 
for Majorana fermions is the anapole from-factor $F_4$ \cite{cpt1, cpt2, cpt3, cpt4, cpt5}.

We next turn our attention to the same electromagnetic matrix element, but now calculated in
$s$-channel (i.e. two Majorana particles outgoing). We write
\begin{eqnarray} \label{schannel}
&&\left(\Omega, j^{\mu}_{em}(0)a^{\dagger}(\vec{k}_1, \lambda_1)a^{\dagger}(\vec{k}_2, \lambda_2)
\Omega \right)
\nonumber \\
&=& \bar{v}(\vec{k}_2, \lambda_2)\Sigma_s^{\mu}(k_1,k_2)u(\vec{k}_1, \lambda_1)
\end{eqnarray}
and assume, in the first instance, that $\Sigma^{\mu}_s$ has exactly the same form-factor
decomposition as in (\ref{formfactor}) with $t$ replaced by $s$. We proceed then along the same 
lines as above. Since the steps in performing the CPT-transformation are very similar to the 
$t$-channel case we only quote the final result. The CPT-transformation gives now
\begin{eqnarray} \label {schannel2}
&& \bar{v}(\vec{k}_2, \lambda_2)\Sigma_s^{\mu}(k_1,k_2)u(\vec{k}_1, \lambda_1)
\nonumber \\
&=&(\lambda^{\ast}_M)^2(-1)^{{\lambda_1 -\lambda_2 \over 2}}
\left(\Omega, j^{\mu}_{em}(0)a^{\dagger}(\vec{k}_1, -\lambda_1)a^{\dagger}(\vec{k}_2, -\lambda_2)
\Omega \right)^{\ast}
\nonumber \\
&=&(\lambda^{\ast}_M)^2(-1)^{{\lambda_1 -\lambda_2 \over 2}} 
\biggl[\bar{v}(\vec{k}_2, -\lambda_2)\Sigma_s^{\mu}(k_1,k_2)u(\vec{k}_1, -\lambda_1)\biggr]^{\dagger}
\end{eqnarray}
As before, we evaluate the last expression for an individual $\Gamma$ matrix and get
\begin{eqnarray} \label{strafo2}
&& (\lambda_M^{\ast})^2(-1)^{{\lambda_1-\lambda_2 \over 2}}
\left[
\bar{v}(\vec{k}_2, -\lambda_2)
\left\{
\begin{array}{c}
 1 \\
i 
\end{array}
\right\}
\Gamma u(\vec{k}_1, -\lambda_1)\right]^{\dagger}
\nonumber \\
&=& (\lambda_M^{\ast})^2\eta_0[\Gamma]\eta_C[\Gamma]\eta_5[\Gamma]\bar{v}(\vec{k}_2, \lambda_2)
\left\{
\begin{array}{c}
 -1  \\
i 
\end{array}
\right\}\Gamma u(\vec{k}_1, \lambda_1)
\end{eqnarray}
It is obvious that with respect to CPT-transformations our conclusion would depend now on the 
{\it choice} of $\lambda_M$! For instance, putting $\lambda_M=\pm i$ seemingly excludes $F_4$
in the $s$-channel. Either CPT would be broken or the Majorana creation is would not be conventional.
Both conclusions are physically not acceptable. The remedy is at hand when
we change equation (\ref{schannel}) by multiplying the right hand side with $\lambda_M^{\ast}$ i.e.
\begin{eqnarray} \label{schannel3}
&&\left(\Omega, j^{\mu}_{em}(0)a^{\dagger}(\vec{k}_1, \lambda_1)a^{\dagger}(\vec{k}_2, \lambda_2)
\Omega \right)
\nonumber \\
&=& \lambda_M^{\ast}\bar{v}(\vec{k}_2, \lambda_2)\Sigma_s^{\mu}(k_1,k_2)u(\vec{k}_1, \lambda_1)
\end{eqnarray}
where $\Sigma_s^{\mu}$ has still the same form-factor decomposition as in (\ref{formfactor}) with
$t$ replaced by $s$. The consequence is now 
that equation (\ref{strafo2}) becomes
\begin{eqnarray} \label{strafo3}
&& (\lambda_M^{\ast})^2(-1)^{{\lambda_1-\lambda_2 \over 2}}
\left[\lambda_M^{\ast}
\bar{v}(\vec{k}_2, -\lambda_2)
\left\{
\begin{array}{c}
 1 \\
i 
\end{array}
\right\}
\Gamma u(\vec{k}_1, -\lambda_1)\right]^{\dagger}
\nonumber \\
&=& \lambda_M^{\ast}\eta_0[\Gamma]\eta_C[\Gamma]\eta_5[\Gamma]\bar{v}(\vec{k}_2, \lambda_2)
\left\{
\begin{array}{c}
 -1  \\
i 
\end{array}
\right\}\Gamma u(\vec{k}_1, \lambda_1)
\end{eqnarray}
which leads to the same results now as in the case of the $t$-channel, namely excluding all 
form-factors except $F_4$. This means, however, also that $s$-channel matrix elements with
Majorana fermions pick up the complex conjugate of the creation phase as an overall phase,
independent of the operator involved. Once the creation phase is fixed, the negligence of this
overall phase in $s$-channel matrix elements leads to disastrous consequences for CPT-properties.
For calculable matrix elements, such as involving a normal product of free Majorana fields,
this overall phase can be justified directly. We get
\begin{eqnarray} \label{direct}
&&\left(\Omega, :\overline{\Psi}_M(x)\Gamma\Psi_M(x):a^{\dagger}(\vec{k}_1,\lambda_1)
a^{\dagger}(\vec{k}_2, \lambda_2)\Omega\right)
\nonumber \\
&=&\lambda_M^{\ast}e^{-i(k_1+k_2)x}(1+\eta_C[\Gamma])\bar{v}(\vec{k}_2,\lambda_2)\Gamma u(\vec{k}_1,
\lambda_1)
\end{eqnarray}
Viewing Feynman rules as Fourier transform of the functional derivatives of the action, it is clear
that the appearance of the global phase in the last two equations is not part of these rules.

In summary, although the creation phase is conventional and unmeasureable, $s$-channel matrix elements
have to be parametrized in the way indicated by equations (\ref{schannel3}) and (\ref{direct}) to avoid
conflict with CPT-invariance.
\vskip 1cm
{\bf Acknowledgments}. 
The author would like to thank the Asian Pacific Center For Theoretical Physics (APCTP, Seoul) for kind 
hospitality and financial support. Useful discussions with G. Cvetic and M. Drees are gratefully
acknowledged.


\vskip 3cm
\begin{thebibliography}{99}
\bibitem{susy}
H. E. Haber and G. L. Kane, Phys. Rep. {\bf 117} (1985) 75; H. P. Nilles, Phys. Rep. {\bf 110}
(1984) 1  
\bibitem{petcov}
S. M. Bilenky and S. T. Petcov, Rev. Mod. Phys. {\bf 59} (1987) 671 
\bibitem{kaiser}
B. Kayser, F. Gibrat-Debu and F. Perrier, ``{\it The Physics of Massive Neutrinos}'', 
World Scientific 1989
\bibitem{mohapatra}
R. N. Mohapatra and P. B. Pal, ``{\it Massive Neutrinos in Physics and Astrophysics}'',
World Scientific 1991;  C. W. Kim and A. Pevsner, ``{\it Neutrinos in Physics and Astrophysics}'',
Harwood 1993 
\bibitem{cpt1}
B. Kayser, Phys. Rev. {\bf D26} (1982) 1662
\bibitem{cpt2}
J. F. Nieves, Phys. Rev. {\bf D26} (1982) 3152
\bibitem{cpt3}
B. Kayser, Phys. Rev. {\bf D28} (1983) 2341
\bibitem{cpt4}
B. Kayser, Phys. Rev. {\bf D30} (1984) 1023
\bibitem{cpt5}
E. E. Radescu, Phys. Rev. {\bf D32} (1985) 1266
\end{thebibliography}
\end{document}